\documentclass[sigconf,natbib=true]{acmart}
\usepackage{graphicx} %
\usepackage[show]{chato-notes}

\newcommand{\micha}[1]{\textcolor{black}{#1}}
\newcommand{\skn}[1]{\textcolor{black}{#1}}

\renewcommand\footnotetextcopyrightpermission[1]{} %
\title{LLMs for estimating positional bias in logged interaction data}

\author{Aleksandr V. Petrov}
\affiliation{
  \institution{Viator, Tripadvisor}
  \city{Glasgow}
  \country{UK}}
\email{firexel@gmail.com}

\author{Michael Murtagh}
\affiliation{
  \institution{Viator, Tripadvisor}
  \city{Lisbon}
  \country{Portugal}}
\email{mmurtagh@tripadvisor.com}

\author{Karthik Nagesh}
\affiliation{
  \institution{Viator, Tripadvisor}
  \city{London}
  \country{UK}}
\email{knagesh@tripadvisor.com}

\usepackage{marginnote}
\newcommand{\pageenlarge}[1]{\enlargethispage{#1\baselineskip}}

\date{June 2025}
\copyrightyear{2025}
\acmYear{2025}
\setcopyright{cc}
\setcctype{by}
\acmConference[CONSEQUENCES@RecSys'25]{Proceedings of the CONSEQUENCES Workshop}{September 22, 2025}{Prague, Czech Republic}

\begin{document}

\begin{abstract}
Recommender and search systems commonly rely on Learning To Rank models trained on logged user interactions to order items by predicted relevance. However, such interaction data is often subject to position bias, as users are more likely to click on items that appear higher in the ranking, regardless of their actual relevance. As a result, newly trained models may inherit and reinforce the biases of prior ranking models rather than genuinely improving relevance.
A standard approach to mitigate position bias is Inverse Propensity Scoring (IPS), where the model’s loss is weighted by the inverse of a propensity function, an estimate of the probability that an item at a given position is examined. However, accurate propensity estimation is challenging, especially in interfaces with complex non-linear layouts.
In this paper, we propose a novel method for estimating position bias using Large Language Models (LLMs) applied to logged user interaction data. This approach offers a cost-effective alternative to online experimentation. Our experiments show that propensities estimated with our LLM-as-a-judge approach are stable across score buckets and reveal the row–column effects of Viator’s grid layout that simpler heuristics overlook. An IPS-weighted reranker trained with these propensities matches the production model on standard NDCG@10 while improving weighted NDCG@10 by roughly 2\%. We will verify these offline gains in forthcoming live-traffic experiments.

\end{abstract}

\maketitle

\pageenlarge{3}
\section{Introduction}
One of the most common setups in modern information retrieval systems, including recommender systems and web search, is the \emph{retrieve-then-rerank} architecture, in which a \emph{candidate generator} model first narrows down the large item catalog to a set of promising candidates, and a reranker model then ranks these candidates according to their estimated relevance to the user~\cite{nogueira2019passage,covington2016deep}.  
According to the Probability Ranking Principle (PRP)~\cite{robertson1977probability}, one of the foundational principles in Information Retrieval, an optimal ranker should order items by their probability of being relevant to the user, $P(\text{relevant} \mid \text{item}, \text{user})$.\footnote{Throughout this paper we use recommender systems terminology; in classical IR, PRP prescribes ranking documents by $P(\text{relevant} \mid \text{query}, \text{document})$.}

In practice, relevance probabilities required by the PRP are typically estimated using machine learning models -- an approach known as Learning to Rank (LTR). Usually LTR models are trained as supervised machine learning models, that rely on labeled data indicating user preferences — for example, clicks, purchases — to learn how to assign higher scores to more relevant items. However, logged interaction data usually suffers from position bias, i.e. users tend to click on higher-ranked products with higher probability. One of the most common models that describes clicks distribution in logged data is the Position-Based Model (PBM, a.k.a. the \micha{Examination Hypothesis})~\cite{craswell2008experimental}, which \micha{assumes that the probability of a click is given by the product of the relevance probability and the positional bias or examination probability}: 
\begin{align}
P&(\text{click}  \mid \text{user}, \text{item}, \text{position}) = \nonumber\\
 &= \underbrace{\colorbox{lightgreen}{ $P(\text{relevant} \mid \text{item}, \text{user})$}}_{\text{relevance probability}}
\;\cdot\;
\underbrace{\colorbox{pink}{$P(\text{examination}\mid\text{position})$}}_{\text{position bias (propensity function)}}
\label{eq:pbm_model}
\end{align}

As evident from Equation~\eqref{eq:pbm_model}, the click probability depends on both item relevance and position bias. Consequently, naïvely training a model via maximum likelihood estimation on raw click logs leads to biased learning, where the model disproportionately favors items that appeared in higher ranks under the previous ranking system. Hence, the model training process \micha{must attempt to mitigate position bias}, for example using Inverse Propensity Scoring (IPS) technique~\cite{joachims2017unbiased}, where the loss function is reweighted by the inverse of the estimated propensity of a position being examined. This results in an \micha{asymptotically} unbiased estimator of relevance, under the assumption that the propensity function $P(\text{examination} \mid \text{position})$ is known or can be reliably estimated.

\begin{figure}
    \centering
    \includegraphics[width=\linewidth]{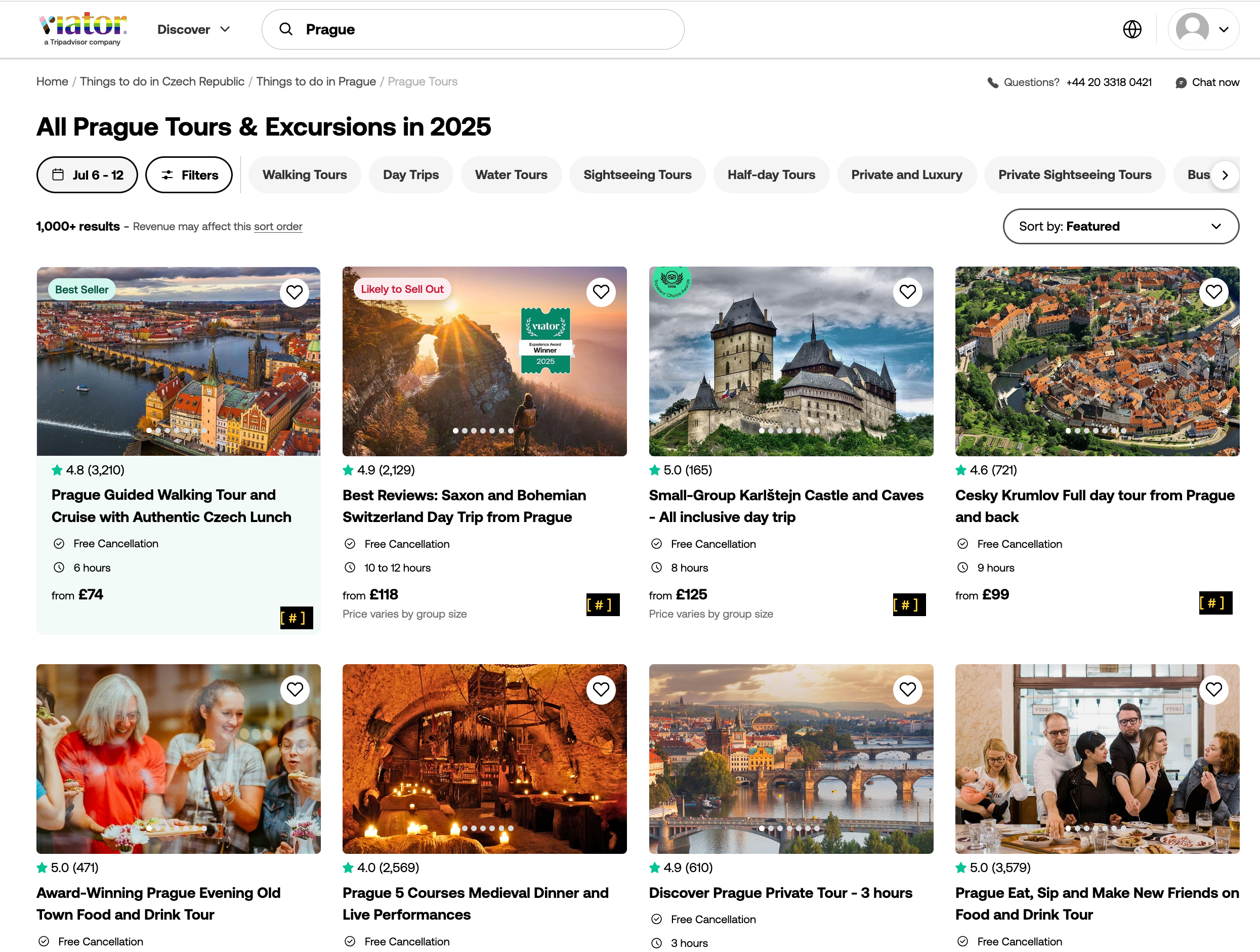}
    \caption{Grid representation \micha{of} search results \micha{on} Viator. With this layout, click propensity may depend on both the row and the column, violating \micha{the} tradional ``linear'' scanning asssumption.}
    \vspace{-1.5\baselineskip}
    \label{fig:grid_layout}
\end{figure}

\pageenlarge{3}
\looseness -1 Hence, accurately estimating the positional bias $P(\text{examination} \mid \text{position})$ is essential for training unbiased LTR models. A common recommendation in the literature is to perform low-impact interventions with real users, such as randomly swapping items in the ranked list, to obtain reliable estimates of this bias~\cite{joachims2017unbiased}. However, in real-world systems, online experimentation is often a scarce and tightly managed resource, with experiment schedules planned months in advance. \skn{Moreover, this approach is constrained by the volume of online traffic: when traffic is limited, allocating a portion for exploration can face (strong) resistance from business stakeholders.} As a result, relying on online interventions to estimate examination bias may not always be feasible.
An alternative is to assume a predefined functional form for the propensity, such as exponential propensity decay~\cite{joachims2017unbiased, ovaisi2020correcting} $P(\text{examination} \mid \text{position}) = \frac{1}{\text{position}^\gamma}$, where $\gamma$ is a tunable hyperparameter. Unfortunately, such assumptions may fail in non-standard interface layouts. For example, Figure~\ref{fig:grid_layout} shows a grid-based search results page used by \skn{Viator.com, one of the largest travel experiences e-commerce platform}. In this layout, the probability that a user examines an item can depend on both the row and the column position, making simple position-based assumptions for the propensity function potentially inaccurate.

This motivates us to explore whether or not it is possible to apply Large Language Models (LLMs) to estimate positional bias in logged click data. Indeed, it has been shown that LLMs are capable to judge relevance of queries to the documents \micha{~\cite{baiduJun2025, zalando2024}} . In the next section, we describe our proposed methodology for estimating propensities using the LLM-as-a-judge approach.
\vspace{-0.5\baselineskip}
\section{Estimating Propensities using an LLM}\label{sec:methodology}

A zero-shot relevance model estimates \micha{relevance} for a user-item pair without being trained on real click logs. Examples include training-free, content-based models such as TF-IDF~\cite{sparck1972statistical} and BM25~\cite{robertson1995okapi}, as well as more recent zero-shot LLM-based relevance estimators, often referred to as LLM-as-a-judge methods~\cite{li2025generation,sun2023chatgpt} \footnote{In this paper, we assume that both user and item are represented using text (e.g. user can be represented using most recent search query and item can be represented using its description).}.  Since zero-shot models are not trained on position-biased click data, we argue that their relevance estimates may be less affected by the logging policy. As such, they can provide a useful signal for propensity estimation, offering judgments that are potentially free from position bias.

\looseness -1 However, we argue that outputs from lexical models such as BM25 are poorly suited for propensity estimation. These models rely solely on lexical overlap, ignoring semantic similarity, and thus often exhibit limited correlation with human relevance \micha{judgements}. Furthermore, lexical similarity is frequently used as an input feature in production ranking models, making the logged item ranks dependent on lexical matching scores. This dependence means lexical signals can indirectly affect position bias, undermining the assumption \micha{of the Position-Based Model} of independence between relevance and examination needed for reliable propensity estimation.

\pageenlarge{3}
\looseness -1In contrast, LLM-as-a-judge methods can provide significantly more accurate relevance estimates, showing strong alignment with human preferences\micha{~\cite{arabzadeh2025benchmarking}}. In our experiments, we evaluated a zero-shot, pointwise\footnote{That is, the model returns a relevance score for each $\langle \text{user}, \text{product} \rangle$ pair.} LLM-as-a-judge model based on GPT-4o-mini. As shown in Table~\ref{tab:llm_ndcg}, this model reduced the NDCG@10 gap\footnote{We do not report absolute metrics for production models in line with company policy.} to the production model by over one-third compared to BM25 (-27.44\% vs. -42.24\%), using Viator bookings as relevance labels.
Moreover, while the LLM-as-a-judge model achieved a lower NDCG@10 than the production model (which was used to generate the logged rankings), this discrepancy is largely attributable to position bias. When evaluating with the weighted version of NDCG (wNDCG)\footnote{\skn{To compute wNDCG, we assign each test instance a weight equal to the position at which the booked product appeared in the logged ranking. While we do not assume access to the true propensity weights, this weighting scheme emphasizes samples where the user likely examined more of the list. As a result, wNDCG is expected to be less affected by position bias than standard NDCG.}}, the LLM-as-a-judge model outperformed the production model. This suggests that the lower NDCG observed is primarily due to the bias introduced by item positions, not inferior relevance estimation.

Currently, LLM relevance estimations are rarely used for ranking in production systems, mainly due to their high inference cost (e.g., in our case exceeding 2 seconds per query-document pair). Consequently, unlike lexical scores (e.g., BM25), LLM-based scores have limited causal influence on item positions in logged data, making them a more viable candidate for use in position-bias–robust propensity estimation.

Our approach for estimating propensities $P(\text{examination} \mid \text{position})$ using LLM-as-a-judge scores is based on two assumptions:\\
\textbf{Assumption 1.}  \emph{A user always examines item at \micha{the} top rank:} 
\begin{align}
    P(\text{examination} \mid \text{position}=1) = 1.0 \label{eq:examination}
\end{align}
Setting $P(\text{examination}\mid\text{position}=1)=1$ is the standard anchor: the vector of propensities is identifiable only up to a constant, so fixing the top slot simply sets the scale (see also the Dynamic Bayesian Network (DBN) model, where the top-ranked item is always examined by setting $\beta_1 = 1$~\cite{chapelle2009dynamic}, and the Dependent Click Model (DCM), which similarly assumes $\mathcal{E}_1 = 1$ for the first rank~\cite{guo2009efficient}). Eye-tracking shows the first result is examined in $\approx$90 \% of desktop sessions \cite{pan2007google, joachims2005accurately}, making the approximation practically harmless.
\pageenlarge{3}

\textbf{Assumption 2.} \emph{Relevance is conditionally independent of position\micha{,} given the LLM-as-a-judge score:}
\begin{align}
    P(\text{relevant} \mid \text{position}, \text{LLM score}) = P(\text{relevant} \mid \text{LLM score})
\end{align}
Assumption~2 implies that, once we condition on the LLM-as-a-judge score, the logged position of an item provides no additional information about its true relevance. This would not hold if we relied on position alone, since positions are determined by the production ranking model, which is explicitly optimised to order items by predicted relevance. Consequently, relevance and position are inherently correlated in logged data unless we control for an independent relevance signal, such as that provided by the LLM-as-a-judge model. This assumption is credible only when the LLM-as-a-judge relevance model is strong and conditions on essentially all relevance-related information available to the production model; otherwise residual dependence may persist. In particular, this restricts the use of simpler lexical models (e.g. BM25) as independent relevance estimators for this purpose.

\begin{table}
    \centering
    \small
    \begin{tabular}{lcc}
        \toprule
        Ranking Method & $\Delta$ NDCG@10 & $\Delta$ wNDCG@10 \\
        \midrule
            BM25 & -42.24\% & -11.20\%  \\ 
            LLM-as-a-judge & \textbf{-27.44\%} &  \textbf{+0.81\%}\\
        \bottomrule
    \end{tabular}
    \caption{Effectiveness of ranking methods on logged Viator bookings, reported as percentage differences ($\Delta$) in NDCG@10 and wNDCG@10 relative to the production model. wNDCG weighs each query by the booked item’s rank in the production model, emphasising harder cases. LLM-as-a-judge assigns pointwise scores via GPT-4o-mini.}
    \vspace{-2\baselineskip}
    \label{tab:llm_ndcg}
\end{table}

Let us consider all logged recommendations shown at position $p$ for which the LLM-as-a-judge system returned a relevance score $s$. In this work, we assume that \micha{the LLM} returns scores from a finite discrete set $\mathcal{S}$; for example in our experiments we instructed an LLM to return integer scores between 0 and 100.  
Given sufficient logged data, we can estimate the click probability $P(\text{click} \mid s, p)$ simply as the proportion of clicks observed in the log. Combining the standard Position-Based Model (Eq.~\eqref{eq:pbm_model}) with Assumption 2 (conditional \micha{independence} of relevance from position) we can factorise click probability for a given LLM Score and positon:

\begin{align}
P(\text{click} \mid s, p) &= P(\text{relevant} \mid s, p) \cdot P(\text{examination} \mid p) \label{eq:prop_for_score}  \nonumber\\ 
&\ \ \ \ \ \ (\text{\small{using Assumption 2}})\\ \nonumber
 &= P(\text{relevant} \mid s)\cdot P(\text{examination} \mid p)
\end{align}
Using Assumption~1, we can estimate the relevance probability for a given LLM-as-a-judge score as:
\begin{align}
P(\text{relevant} \mid s) = P&(\text{click} \mid s, \text{position} = 1) \label{eq:rel_estimation}
\end{align}
In other words, \emph{the probability that an item is relevant to a user given an LLM-as-a-judge score $s$ can be estimated as the proportion of clicks among user–item pairs with the same LLM score shown at the top position.} 

Finally, subsituting Equation~\eqref{eq:rel_estimation} into Equation.~\eqref{eq:prop_for_score}, we obtain propensity estimation for any position $p$:
\begin{equation}
    P(\text{examination} \mid position=p) = \frac{P(\text{click} \mid \text{} s, position=p)}{P(\text{click} \mid s, position=1)}  
\end{equation}
where click probabilities \micha{on} the right are obtained from click logs for some specific value of \micha{the} LLM-as-a-judge score $s$. 

While we can technically choose any  $s \in \mathcal{S}$ and only use it for estimating propensities, \micha{doing so would} ignore \micha{large amounts} of \micha{the} available data, potentially making our \micha{estimates} prone to noise. Hence, to make our  \micha{estimates} more robust, we propose to average results across all possible values of LLM-as-a-judge scores $s$: 
\begin{equation}
    P(\text{examination} \mid position=p) = \frac{1}{|\mathcal{S}|}\sum_{s\in\mathcal{S}}\frac{P(\text{click} \mid s, position=p)}{P(\text{click} \mid {s}, position=1)}  \label{eq:main_eq}
\end{equation}
\pageenlarge{3}
In the next section, we also explore how restricting the averaging to specific subsets of $\mathcal{S}$, e.g., using only high-relevance or low-relevance scores, affects the resulting estimates. This allows us to assess the sensitivity of our method to different relevance levels inferred by the LLM.

\begin{figure}[tb]
    \centering
    \includegraphics[width=0.75\linewidth]{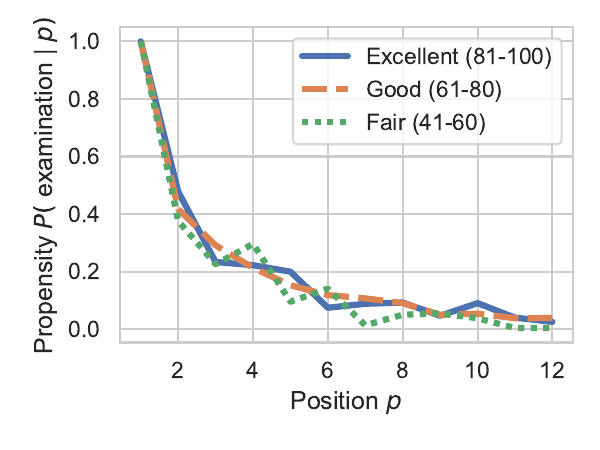}
    \vspace{-1\baselineskip}
   \caption{Recovered propensity scores using different levels of LLM Relevance judgements.}
   \label{fig:props_by_score}
   \vspace{-1.5\baselineskip}
\end{figure}

\vspace{-0.5\baselineskip}
\section{Experiments \& Discussion}
In this section we present \micha{our} experimental analysis of propensity score estimation. We use an LLM-as-a-judge-based propensity estimation approach to estimate click propensities on Viator’s search results page. Specifically, we \micha{randomly} sample 8,439 logged search \micha{queries and the corresponding search results page impressions} \micha{between March and May 2025 worth of user logs}. For each query, we assign a relevance score between 0 and 100 using the GPT-4o-mini model. We then apply the methodology described in Section~\ref{sec:methodology}, in particular, Equation~\eqref{eq:main_eq}, to recover position propensities for Viator’s grid-based search layout. In our experiments, we analyse two research questions, which \micha{are now described in turn}.

\noindent\textbf{RQ1. How sensitive are the estimated examination propensities to the choice of LLM‐as‐a‐judge score bucket?}

Equation~\eqref{eq:main_eq} implies that \emph{if} Assumption 2 (conditional independence of relevance and position given the LLM score) holds, averaging over any subset $\mathcal{S}\subseteq[0,100]$ should leave the \emph{shape} of the propensity curve unchanged, only its sampling variance should differ.  If the assumption is violated, however, residual relevance–position correlations may persist within particular score ranges, yielding bucket-specific propensity curves.  Comparing these curves therefore provides a diagnostic of how well Assumption 2 holds in practice.

To probe this, we partition the LLM scores into three buckets—\emph{Excellent} (81–100), \emph{Good} (61–80), and \emph{Fair} (41–60).  Scores below 41 are too sparse to analyse: the production ranker’s guardrails already suppress items that the LLM deems clearly irrelevant.  We then estimate propensity curves separately for each bucket (Figure~\ref{fig:props_by_score}).  The three curves are virtually superimposed, exhibiting no systematic divergence across ranking positions.  Hence the propensity estimator appears robust to the choice of score bucket, providing empirical support for the conditional-independence assumption and suggesting that the learned propensities are close to the true examination probabilities.

\noindent \textbf{RQ2. How does grid layout affect examination propensities?}

\begin{figure}[tb]
    \centering
    \includegraphics[width=0.65\linewidth]{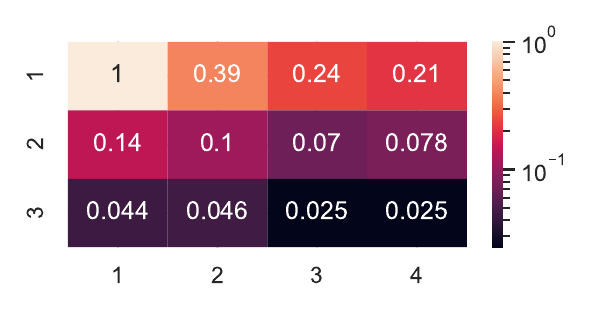}
    \vspace{-1\baselineskip}
   \caption{Estimated examination propensities, $P(\text{examination}\mid\text{position})$, for the first three rows of Viator’s search-results grid layout.}
       \vspace{-1\baselineskip}
    \label{fig:probs_on_grid}
\end{figure}

We apply Equation~\eqref{eq:main_eq} to the full set of LLM‐as‐a‐judge scores and mapped the resulting propensities onto Viator’s grid (Figure~\ref{fig:probs_on_grid}).  The top-left cell unsurprisingly attracts the highest examination probability, yet several layout-specific deviations emerge.  The drop from the first to the second column is markedly steeper than the drop from the third to the fourth, and in the second row the propensity even \emph{rises} by roughly 10 \% from the third to the fourth cell, likely because the right-edge item anchors at the screen border while its neighbour is visually “lost in the middle.”  In the third row, which usually becomes visible only after a short scroll, the left–right gap nearly vanishes, indicating that lateral bias weakens once users scroll.  Hence, the grid induces row- and column-dependent effects that a one-dimensional propensity model cannot capture.

\pageenlarge{3}
\vspace{-0.5\baselineskip}
\section{Conclusion}
We introduced an LLM-as-a-judge method for offline estimation of examination propensities in grid layouts.  Score-conditioned curves overlap, supporting our conditional-independence assumption, while the learned propensities expose clear row–column effects that one-dimensional heuristics miss.  In our early experiments,  IPS-weighted reranker trained with these propensities performed on par \micha{with} the production model on NDCG@10 but showed better results on weighted NDCG(+2\%). We are planning live experiments to validate these gains.

\bibliographystyle{ACM-Reference-Format}
\balance
\bibliography{references}


\begin{thebibliography}{17}


\ifx \showCODEN    \undefined \def \showCODEN     #1{\unskip}     \fi
\ifx \showDOI      \undefined \def \showDOI       #1{#1}\fi
\ifx \showISBNx    \undefined \def \showISBNx     #1{\unskip}     \fi
\ifx \showISBNxiii \undefined \def \showISBNxiii  #1{\unskip}     \fi
\ifx \showISSN     \undefined \def \showISSN      #1{\unskip}     \fi
\ifx \showLCCN     \undefined \def \showLCCN      #1{\unskip}     \fi
\ifx \shownote     \undefined \def \shownote      #1{#1}          \fi
\ifx \showarticletitle \undefined \def \showarticletitle #1{#1}   \fi
\ifx \showURL      \undefined \def \showURL       {\relax}        \fi
\providecommand\bibfield[2]{#2}
\providecommand\bibinfo[2]{#2}
\providecommand\natexlab[1]{#1}
\providecommand\showeprint[2][]{arXiv:#2}

\bibitem[Arabzadeh and Clarke(2025)]%
        {arabzadeh2025benchmarking}
\bibfield{author}{\bibinfo{person}{Negar Arabzadeh} {and}
  \bibinfo{person}{Charles~LA Clarke}.} \bibinfo{year}{2025}\natexlab{}.
\newblock \showarticletitle{Benchmarking LLM-based relevance judgment methods}.
  In \bibinfo{booktitle}{\emph{Proceedings of the 48th International ACM SIGIR
  Conference on Research and Development in Information Retrieval}}.
  \bibinfo{pages}{3194--3204}.
\newblock


\bibitem[Chapelle and Zhang(2009)]%
        {chapelle2009dynamic}
\bibfield{author}{\bibinfo{person}{Olivier Chapelle} {and} \bibinfo{person}{Ya
  Zhang}.} \bibinfo{year}{2009}\natexlab{}.
\newblock \showarticletitle{A dynamic bayesian network click model for web
  search ranking}. In \bibinfo{booktitle}{\emph{Proceedings of the 18th
  international conference on World wide web}}. \bibinfo{pages}{1--10}.
\newblock


\bibitem[Covington et~al\mbox{.}(2016)]%
        {covington2016deep}
\bibfield{author}{\bibinfo{person}{Paul Covington}, \bibinfo{person}{Jay
  Adams}, {and} \bibinfo{person}{Emre Sargin}.}
  \bibinfo{year}{2016}\natexlab{}.
\newblock \showarticletitle{Deep neural networks for youtube recommendations}.
  In \bibinfo{booktitle}{\emph{Proceedings of the 10th ACM conference on
  recommender systems}}. \bibinfo{pages}{191--198}.
\newblock


\bibitem[Craswell et~al\mbox{.}(2008)]%
        {craswell2008experimental}
\bibfield{author}{\bibinfo{person}{Nick Craswell}, \bibinfo{person}{Onno
  Zoeter}, \bibinfo{person}{Michael Taylor}, {and} \bibinfo{person}{Bill
  Ramsey}.} \bibinfo{year}{2008}\natexlab{}.
\newblock \showarticletitle{An experimental comparison of click position-bias
  models}. In \bibinfo{booktitle}{\emph{Proceedings of the 2008 international
  conference on web search and data mining}}. \bibinfo{pages}{87--94}.
\newblock


\bibitem[Guo et~al\mbox{.}(2009)]%
        {guo2009efficient}
\bibfield{author}{\bibinfo{person}{Fan Guo}, \bibinfo{person}{Chao Liu}, {and}
  \bibinfo{person}{Yi~Min Wang}.} \bibinfo{year}{2009}\natexlab{}.
\newblock \showarticletitle{Efficient multiple-click models in web search}. In
  \bibinfo{booktitle}{\emph{Proceedings of the second acm international
  conference on web search and data mining}}. \bibinfo{pages}{124--131}.
\newblock


\bibitem[Hosseini et~al\mbox{.}(2025)]%
        {zalando2024}
\bibfield{author}{\bibinfo{person}{Kasra Hosseini}, \bibinfo{person}{Thomas
  Kober}, \bibinfo{person}{Josip Krapac}, \bibinfo{person}{Roland Vollgraf},
  \bibinfo{person}{Weiwei Cheng}, {and} \bibinfo{person}{Ana
  Peleteiro~Ramallo}.} \bibinfo{year}{2025}\natexlab{}.
\newblock \showarticletitle{Retrieve, Annotate, Evaluate, Repeat: Leveraging
  Multimodal LLMs for Large-Scale Product Retrieval Evaluation}. In
  \bibinfo{booktitle}{\emph{European Conference on Information Retrieval}}.
  Springer, \bibinfo{pages}{149--163}.
\newblock


\bibitem[Joachims et~al\mbox{.}(2005)]%
        {joachims2005accurately}
\bibfield{author}{\bibinfo{person}{Thorsten Joachims}, \bibinfo{person}{Laura
  Granka}, \bibinfo{person}{Bing Pan}, \bibinfo{person}{Helene Hembrooke},
  {and} \bibinfo{person}{Geri Gay}.} \bibinfo{year}{2005}\natexlab{}.
\newblock \showarticletitle{Accurately interpreting clickthrough data as
  implicit feedback}. In \bibinfo{booktitle}{\emph{Proceedings of the 28th
  Annual International ACM SIGIR Conference on Research and Development in
  Information Retrieval}} (Salvador, Brazil) \emph{(\bibinfo{series}{SIGIR
  '05})}. \bibinfo{publisher}{Association for Computing Machinery},
  \bibinfo{address}{New York, NY, USA}, \bibinfo{pages}{154–161}.
\newblock
\showISBNx{1595930345}
\urldef\tempurl%
\url{https://doi.org/10.1145/1076034.1076063}
\showDOI{\tempurl}


\bibitem[Joachims et~al\mbox{.}(2017)]%
        {joachims2017unbiased}
\bibfield{author}{\bibinfo{person}{Thorsten Joachims}, \bibinfo{person}{Adith
  Swaminathan}, {and} \bibinfo{person}{Tobias Schnabel}.}
  \bibinfo{year}{2017}\natexlab{}.
\newblock \showarticletitle{Unbiased learning-to-rank with biased feedback}. In
  \bibinfo{booktitle}{\emph{Proceedings of the tenth ACM international
  conference on web search and data mining}}. \bibinfo{pages}{781--789}.
\newblock


\bibitem[Li et~al\mbox{.}(2025)]%
        {li2025generation}
\bibfield{author}{\bibinfo{person}{Dawei Li}, \bibinfo{person}{Bohan Jiang},
  \bibinfo{person}{Liangjie Huang}, \bibinfo{person}{Alimohammad Beigi},
  \bibinfo{person}{Chengshuai Zhao}, \bibinfo{person}{Zhen Tan},
  \bibinfo{person}{Amrita Bhattacharjee}, \bibinfo{person}{Yuxuan Jiang},
  \bibinfo{person}{Canyu Chen}, \bibinfo{person}{Tianhao Wu}, {et~al\mbox{.}}}
  \bibinfo{year}{2025}\natexlab{}.
\newblock \showarticletitle{From generation to judgment: Opportunities and
  challenges of llm-as-a-judge, 2025}.
\newblock \bibinfo{journal}{\emph{URL https://arxiv. org/abs/2411.16594}}
  (\bibinfo{year}{2025}).
\newblock


\bibitem[Nogueira and Cho(2019)]%
        {nogueira2019passage}
\bibfield{author}{\bibinfo{person}{Rodrigo Nogueira} {and}
  \bibinfo{person}{Kyunghyun Cho}.} \bibinfo{year}{2019}\natexlab{}.
\newblock \showarticletitle{Passage Re-ranking with BERT}.
\newblock \bibinfo{journal}{\emph{arXiv preprint arXiv:1901.04085}}
  (\bibinfo{year}{2019}).
\newblock


\bibitem[Ovaisi et~al\mbox{.}(2020)]%
        {ovaisi2020correcting}
\bibfield{author}{\bibinfo{person}{Zohreh Ovaisi}, \bibinfo{person}{Ragib
  Ahsan}, \bibinfo{person}{Yifan Zhang}, \bibinfo{person}{Kathryn Vasilaky},
  {and} \bibinfo{person}{Elena Zheleva}.} \bibinfo{year}{2020}\natexlab{}.
\newblock \showarticletitle{Correcting for selection bias in learning-to-rank
  systems}. In \bibinfo{booktitle}{\emph{Proceedings of The Web Conference
  2020}}. \bibinfo{pages}{1863--1873}.
\newblock


\bibitem[Pan et~al\mbox{.}(2007)]%
        {pan2007google}
\bibfield{author}{\bibinfo{person}{Bing Pan}, \bibinfo{person}{Helene
  Hembrooke}, \bibinfo{person}{Thorsten Joachims}, \bibinfo{person}{Lori
  Lorigo}, \bibinfo{person}{Geri Gay}, {and} \bibinfo{person}{Laura Granka}.}
  \bibinfo{year}{2007}\natexlab{}.
\newblock \showarticletitle{In Google we trust: Users’ decisions on rank,
  position, and relevance}.
\newblock \bibinfo{journal}{\emph{Journal of computer-mediated communication}}
  \bibinfo{volume}{12}, \bibinfo{number}{3} (\bibinfo{year}{2007}),
  \bibinfo{pages}{801--823}.
\newblock


\bibitem[Robertson(1977)]%
        {robertson1977probability}
\bibfield{author}{\bibinfo{person}{Stephen~E Robertson}.}
  \bibinfo{year}{1977}\natexlab{}.
\newblock \showarticletitle{The probability ranking principle in IR}.
\newblock \bibinfo{journal}{\emph{Journal of documentation}}
  \bibinfo{volume}{33}, \bibinfo{number}{4} (\bibinfo{year}{1977}),
  \bibinfo{pages}{294--304}.
\newblock


\bibitem[Robertson et~al\mbox{.}(1995)]%
        {robertson1995okapi}
\bibfield{author}{\bibinfo{person}{Stephen~E Robertson}, \bibinfo{person}{Steve
  Walker}, \bibinfo{person}{Susan Jones}, \bibinfo{person}{Micheline~M
  Hancock-Beaulieu}, \bibinfo{person}{Mike Gatford}, {et~al\mbox{.}}}
  \bibinfo{year}{1995}\natexlab{}.
\newblock \showarticletitle{Okapi at TREC-3}.
\newblock  (\bibinfo{year}{1995}).
\newblock


\bibitem[Sparck~Jones(1972)]%
        {sparck1972statistical}
\bibfield{author}{\bibinfo{person}{Karen Sparck~Jones}.}
  \bibinfo{year}{1972}\natexlab{}.
\newblock \showarticletitle{A statistical interpretation of term specificity
  and its application in retrieval}.
\newblock \bibinfo{journal}{\emph{Journal of documentation}}
  \bibinfo{volume}{28}, \bibinfo{number}{1} (\bibinfo{year}{1972}),
  \bibinfo{pages}{11--21}.
\newblock


\bibitem[Sun et~al\mbox{.}(2023)]%
        {sun2023chatgpt}
\bibfield{author}{\bibinfo{person}{Weiwei Sun}, \bibinfo{person}{Lingyong Yan},
  \bibinfo{person}{Xinyu Ma}, \bibinfo{person}{Shuaiqiang Wang},
  \bibinfo{person}{Pengjie Ren}, \bibinfo{person}{Zhumin Chen},
  \bibinfo{person}{Dawei Yin}, {and} \bibinfo{person}{Zhaochun Ren}.}
  \bibinfo{year}{2023}\natexlab{}.
\newblock \showarticletitle{Is ChatGPT good at search? investigating large
  language models as re-ranking agents}.
\newblock \bibinfo{journal}{\emph{arXiv preprint arXiv:2304.09542}}
  (\bibinfo{year}{2023}).
\newblock


\bibitem[Wang et~al\mbox{.}(2025)]%
        {baiduJun2025}
\bibfield{author}{\bibinfo{person}{Xingzhu Wang}, \bibinfo{person}{Erhan
  Zhang}, \bibinfo{person}{Yiqun Chen}, \bibinfo{person}{Jinghan Xuan},
  \bibinfo{person}{Yucheng Hou}, \bibinfo{person}{Yitong Xu},
  \bibinfo{person}{Ying Nie}, \bibinfo{person}{Shuaiqiang Wang},
  \bibinfo{person}{Dawei Yin}, {and} \bibinfo{person}{Jiaxin Mao}.}
  \bibinfo{year}{2025}\natexlab{}.
\newblock \showarticletitle{Leveraging LLMs to Evaluate Usefulness of
  Document}.
\newblock \bibinfo{journal}{\emph{arXiv preprint arXiv:2506.08626}}
  (\bibinfo{year}{2025}).
\newblock


\end{thebibliography}

\end{document}